\documentstyle[prl,aps,twocolumn,epsf]{revtex}
\def\break#1{\pagebreak \vspace*{#1}}
\def\beq{\begin{eqnarray}}
\def\eeq{\end{eqnarray}}
\begin{document}
\draft
\title{Absence of local thermal equilibrium  in 
two models of heat conduction}  
\author{Abhishek Dhar $^1$ and Deepak Dhar $^2$}
\address{Theoretical Physics Group, Tata Institute of Fundamental Research,
Homi Bhabha Road, Bombay 400005, India.\\
$^1$ e-mail : abhi@theory.tifr.res.in
$^2$ e-mail : ddhar@theory.tifr.res.in }
\date{\today}
\maketitle
\widetext
\begin{abstract}

A crucial assumption in the conventional description of thermal conduction
is the existence of local thermal equilibrium. We test this assumption in
two simple models of heat conduction. Our first model is a linear chain of
planar spins with nearest neighbour couplings, and the second model is
that of a Lorentz gas. We look at the steady state of the system when the
two ends are connected to heat baths at temperatures $T_1$ and $T_2$. If
$T_1 = T_2$, the system reaches thermal equilibrium. If $ T_1 \neq T_2$,
there is a heat current through the system, but there is no local thermal
equilibrium. This is true even in the limit of large system size, when the
heat current goes to zero. We argue that this is due to the existence of
an infinity of local conservation laws in their dynamics.

\end{abstract}

\pacs{PACS numbers: 05.70.Ln, 05.20.-y, 44.10.+i, 75.10.Hk }

\narrowtext

Heat conduction provides a rather elementary example of nonequilibrium
steady states (NSS). Here the driving force on the system is an
externally applied temperature gradient. Our present understanding of
this in terms of nonequilibrium statistical mechanics is not entirely
satisfactory. In the traditional treatment of heat conduction, a very
important assumption is of a rather quick ( in atomic time scales)
establishment of local thermodynamical equilibrium (LTE)  \cite{kreu}.  
This allows one to define thermodynamical quantities such as temperature,
pressure etc. locally. One then writes evolution equations for the slow
change in time in these quantities in terms of their small spatial
gradients. Thus, in heat conduction, LTE implies that we can define a
temperature field $T(x)$, that varies slowly in space. In linear response
theory, the local heat current density $\bar J(x)$ is given by $\bar J=-K(T)
\bar{\nabla} T$ (Fourier's law).  The well-known Kubo formula determines
the thermal conductivity coefficient $K(T)$ in terms of the time-dependent
{\it equilibrium} correlations of the system \cite{kubo}.

However, the question of the necessary (or sufficient) conditions for the
fast establishment or non-establishment of LTE has not been investigated
much so far. This is what we study in this paper, in two simple models.  
The first is a one-dimensional spin model evolving with a Markovian
stochastic dynamics. The second model is that of a $d$-dimensional ($d>1$)
Lorentz gas of 
noninteracting particles scattering elastically with a set of randomly
placed obstacles ({\it e.g} spheres) of finite density. In both these
models, we find the unexpected result that 
the NSS does not show LTE, even in the limit of very small thermal
gradients. Thus, one cannot define a local temperature, and the
conventional linear response theory does not work. We argue that a
sufficient condition for LTE not to occur is the existence of an infinity
of locally conserved quantities in the dynamics.
\break{1.5in}

It is obvious that existence of extra conservation laws in an isolated
classical mechanical system implies breakdown of ergodicity, and hence
failure to reach thermodynamical equilibrium. Our results are nontrivial
in that we show that this occurs {\it even in the presence of coupling to
heat baths} which leads to breakdown of the conservation laws at the
boundaries of the system. Also, the role of conservation laws in
stochastically evolving system is somewhat different from that in
deterministic evolution \cite{menon}, and not so well-studied.

Several microscopic models of heat conduction have been studied in the
past. The simple model of a harmonic lattice, evolving with classical
mechanical equations of motion, and coupled to two heat baths at different
temperatures $T_1$ and $T_2$ at opposite ends of the lattice, is known to
show anomalous conduction \cite{ried}. It is found that the heat current
$J$ across the sample remains finite as the size of the system $L$ goes to
infinity, for a fixed $T_1$ and $T_2$. Simulations of 
harmonic chains with disorder show that $J$ decreases to zero for large
$L$, but only as $L^{-1/2}$ \cite{rubin}. Several models with
nonlinear couplings have been studied numerically: 
the FPU chain with quartic potential \cite{FPU}, the Toda lattice
\cite{toda}, the so-called ding-a-ling model \cite{ding}, the
Frenkel-Kontorova model \cite{bambihu} etc.  
Most of these models are one-dimensional, and all do not show the expected $J
\sim (T_1 - T_2)/L$ behavior. The uniqueness of the steady state has
been proved only for some special models of heat baths, and only if
the temperature difference is sufficiently small \cite{eckmann}. In this
respect models with
stochastic
dynamics have been more successful. These typically work with local
energy-conserving moves, and often involve introduction of additional
degrees of freedom. Creutz has used an algorithm with Maxwell demons to
simulate heat conduction in the Ising model \cite{creutz}. Lebowitz
and Spohn studied heat conduction in the Lorentz model, 
and showed that in the 
Boltzmann-Grad limit of large number of scatterers of very small size,
one recovers the Fourier law ( $ J \sim (T_1 - T_2)/L$) \cite{lebowitz}.

We start with a precise definition of our first model.
We consider a linear chain of $L$ planar spins. The spin at
site $i$ ($1 \leq i \leq L$) of the lattice is specified by the angle
$\theta_i$, $0 \leq
\theta \leq 2\pi$. The spins interact with nearest neighbors by
ferromagnetic coupling $K$. The Hamiltonian of the system is given by

\begin{eqnarray}
{\cal{H}}= - K\sum_{<i,j>} \cos{(\theta_i-\theta_j)} 
\end{eqnarray}
where the sum is over all nearest neighbors.

The dynamics is the following: Suppose the instantaneous local field
at a site $i$ due to coupling of the neighbors is in the direction
$\phi_i$, and the spin at the site is $\theta_i = \phi_i + \delta
\theta_i$. Then the transition 
$\delta \theta_i \rightarrow -\delta\theta_i$
does not change the energy of the system. We assume that such spin flips
occur at all sites stochastically with a constant rate (which may be
chosen to be $1$).
The boundary spins $i= 1,L$ are connected to heat baths.
For these spins, the
flip rates are the following:
A boundary spin $\theta_i$ can change to any value $\theta_i'$ with a rate
$\alpha$ if the energy change $\triangle E \le 0$, and with a rate   
$\alpha e^{-\triangle E /T}$, if $\triangle E>0$; 
where $T$ is the temperature of the heat bath.

Note that in this model energy is conserved exactly away from the
boundaries of the lattice.  In the absence of any coupling to heat baths,
this dynamics has been studied by one of us earlier \cite{abhi}. It was
found that, in general, the dynamics is not ergodic, and the phase space
breaks up into disconnected sectors. We now show that the coupling to
heat bath allows transitions between these sectors, and makes the
system fully ergodic.

Since the rate matrix in our problem satisfies the detailed balance
condition, it is sufficient to show that all configurations can be reached
from any initial configuration. Consider first the case when a single spin
on the lattice, say $\theta_0$, is coupled to a heat bath at temperature
$T$. Then, $\theta_0$ can be made to take any value, since it can exchange
energy with the bath, keeping other spins unchanged. Next consider any spin,
$\theta_1$ that is nearest neighbour to $\theta_0$. Then it is easy to see
that flipping $\theta_1$, then changing $\theta_0$ by a small amount
$\delta \theta_0$, then flipping $\theta_1$ again, then changing
$\theta_0$ back again to its initial value keeps all spins unchanged,
except for $\theta_1$ which changes by an amount proportional to $\delta
\theta_0$.  

We can use a similar argument to change a neighbour of $\theta_1$, and so
on. Thus, there exists an allowed sequence of spin flips by which we can
rotate any single spin by an infinitesimal amount. By making many such
rotations we can reach any spin configuration. This completes the proof of
ergodicity. Note that the proof is valid in all dimensions $d$ and
also if more than one spin is in contact with the same heat bath (
same $T$). 

In one dimension the equilibrium properties of the $XY$ model are easily
determined. We define $\Delta \theta_i = \theta_{i+1} - \theta_i$. Then
clearly the $\Delta \theta_i$ are independent random variables. For a
mesoscopic description, we
define coarse-grained densities by averaging the corresponding
microscopic quantity over length scales $\ell$, with $1 << \ell <<L$. Let
$u^{(n)}(x)$ be the coarse-grained density corresponding to 
$\cos ^n(\Delta \theta_i)$ for $i$ lying in a neighborhood of the
point $x$. In
equilibrium, at inverse temperature $\beta$, there is no dependence on
$x$, and one gets
\begin{eqnarray}
u_i^{(n)}=\frac{\int {\cos ^n(\Delta \theta)} e^{\beta K \cos(\Delta
\theta)} d(\Delta \theta) }{\int e^{\beta K \cos(\Delta \theta)}
d(\Delta \theta) } 
\end{eqnarray}
Eliminating $\beta K$ from these equations we can express $u^{(n)}$, for
$n \ge 2$, as explicit nonlinear functions of $u^{(1)}$, {\it i.e.}
$u^{(n)}=F^{(n)}(u^{(1)})$.

We consider now the case when the two ends are connected to different heat
baths, at temperatures $T_1$ and $T_2$. 
We show that, in the NSS, LTE is not achieved. 

In this dynamics, whenever a spin flip occurs, the values of $\Delta
\theta$ on the two adjacent bonds get interchanged. This means that
the densities $u^{(n)}$ are locally conserved for all positive
integers $n$. For each of these
quantities one can write a corresponding current. The current from the
($x-1$)th bond to the $x$th bond is given by

\begin{eqnarray}
J^{(n)}(x) &=& (-<\cos^n(\Delta \theta_{x-1})>+<\cos^n(\Delta
\theta_{x})>) \nonumber \\
&=& -\nabla u^{(n)}(x)
\end{eqnarray}

In the steady state, in one dimension, $J^{(n)}(x)$ must be independent of
$x$, and hence $u^{(n)}$ vary linearly across the chain for all $n$.
Their value at the two ends are determined by the temperatures at the
ends. Then, eliminating the coordinate $x$, $u^{(n)}(x)$ is expressible as
a linear function of $u^{(1)}(x)$, for all $n$. But, as shown above, {\it
in equilibrium}, $u^{(n)}$ , for $ n \geq 2$ are nonlinear functions of
$u^{(1)}$.  Thus, we have constructed a simple model in which heat current
in the steady state is locally proportional to the gradient of energy
density, but there is no LTE (Fig. 1). This result is true even in the
limit of system size $L \to \infty$, when the heat current through the
lattice becomes infinitesimal.

For a simple extension of the above result, consider a linear chain in
which there is one bond with a different bond-strength $K'$. Then the
quantities $u^{(n)}$ ($n \ge 2$) are not conserved at this `bad'
bond. In this case, 
the currents take different values $J^{(n)}_1$ and $J^{(n)}_2$ 
to the left and right of the bond respectively (for $n \ge 2$). In
this case, there are no strict 
conservation laws, but $ u^{(n)}$ are still linear functions of $u^{(1)}$
in each (left or right) half of the chain, and LTE is still not
obtained. This conclusion has also been checked in simulations (Fig. 2).

In higher dimensions, there are no known locally conserved quantities,
other than energy, in the spin model. Hence, we expect LTE to be attained.
Indeed, a numerical simulation of a $40 \times 40$ lattice with
$T_1=1.4 K$, and 
$T_2=0.4 K$ agrees fully with this conclusion. In this case, we checked
that the observed value of $<u^{(2)}(x)>$ at any point $x$ in the NSS
agrees very well with the value of $u^{(2)}$ in the {\it homogeneous
equilibrium state} having energy density $u^{(1)}(x)$.
Similarly, if we consider a dimerized chain with alternating
coupling constants $K$ and $K'$, then there are no conserved quantities 
other than the energy density. In this case, the results of our
simulations ( also shown in Fig. 1) confirm that the
NSS does show LTE even in one dimension if $K \neq K'$. 

Our second example of absence of LTE in heat conduction is provided by the
model of the $d$-dimensional Lorentz gas studied earlier by Lebowitz and Spohn
\cite{lebowitz}. The model describes a gas of non-interacting point
particles moving in a in box, undergoing elastic scattering by a random
assembly of fixed obstacles of arbitrary shape (Fig. 3). In this
model, all collisions with the obstacles are elastic, and energy is
conserved. But 
collisions between the particles and the two walls (at $x=0$ and
$x=L$) are inelastic, and
lead to the energy after collision being thermalized corresponding to
the temperature of the wall.

Let $\rho(E,x)  dE$ be the average density of particles having energy
between $E$ and $E + dE$ in a small volume centered at point $x$.
Then all moments $\mu^{(n)}(x)=\int E^n \rho(E,x) dE$ of this
distribution function are locally conserved. 

Now consider the effect of coupling it to two different reservoirs
at temperatures $T_1$ and $T_2$. 
Let $p(x)$ be the probability that a randomly chosen particle in the small
volume chosen near $x$ was introduced at the left end. Then as collisions
with scatterers do not change the energies, clearly $\rho(E,x)$ is given by
\begin{eqnarray}
\rho(E,x) = p(x) \rho(E,x=0) + ( 1 - p(x)) \rho(E,x=L)
\end{eqnarray}

The $x$-dependence of $\rho(E,x)$ comes only from the spatial dependence of
$p(x)$. As the linear combination of two maxwellians is not a maxwellian,
and the distribution must be a maxwellian {\it in thermal equilibrium}, it
follows that there is no thermal equilibrium in the Lorentz model
\cite{footnote}. Also we expect the heat flow to be diffusive ( $J
\sim 1/L$) in the
limit $L >> \ell$, where $\ell$ is the mean free path of the Lorentz
particles. Note that an equation similar to Eq. (4) can be 
written down for the first model also.

If we allow inelastic scattering with local energy conservation ( each
scatterer can store a small amount of energy, which can be exchanged with
the scattered particles ), the infinity of conservation laws goes away, and
we expect that then the model would show LTE, and normal conduction.

 To summarize, in this Letter, we studied two simple models of heat
conduction, and showed that local thermal equilibrium is not reached as
both the models have an infinity of locally conserved quantities. These
counter-examples should help understand better the mechanism of local
thermal equilibriation in nonequilibrium systems, in general.

We thank Mohit Randeria and Jean-Pierre Eckmann for critically reading
the manuscript, and one of the referees for several useful comments on an
earlier version of this paper.

%\centerline{\bf Figure Captions}

\vbox{
\epsfxsize=8.0cm
\epsfysize=8.0cm
\epsffile{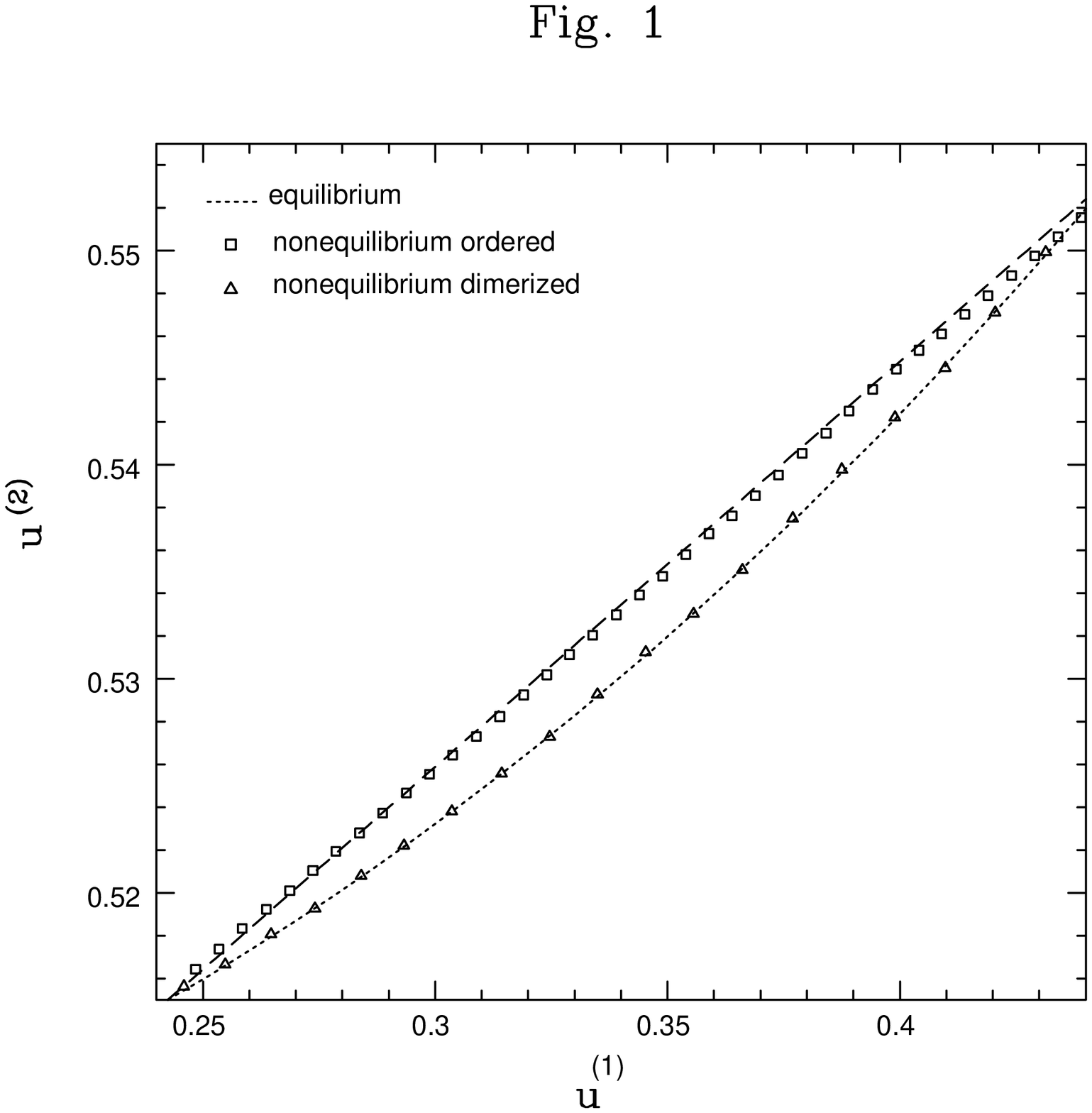}
\begin{figure}
\caption{
Plot of  $u^{(2)}(x)$ versus $u^{(1)}(x)$ in the steady state of the
linear chain ($K=1$). The results for the dimerized chain ($K'=2$)
agree with the equilibrium curve.} 
\label{dimer}
\end{figure}}
\vbox{
\epsfxsize=8.0cm
\epsfysize=8.0cm
\epsffile{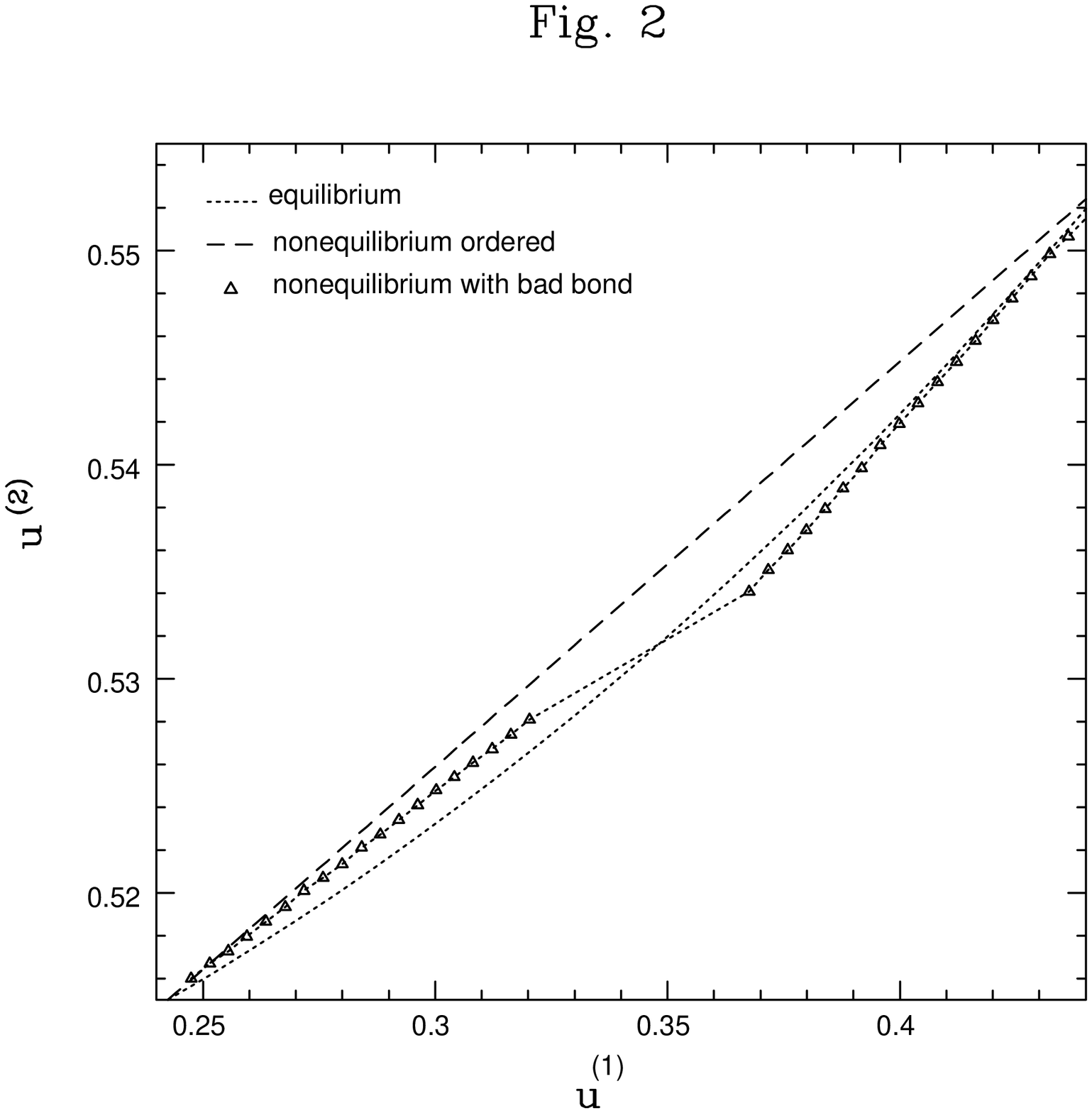}
\begin{figure}
\caption{\label{} 
Plot of $u^{(2)}$ versus $u^{(1)}$ when the middle bond has a different
strength $K'=0.5$ showing the different linear dependences in
the left and right halves of the chain. For comparison, the case
$K'=1$, and the equilibrium curve are also shown.}
\end{figure}}

\vbox{
\epsfxsize=8.0cm
\epsfysize=6.0cm
\epsffile{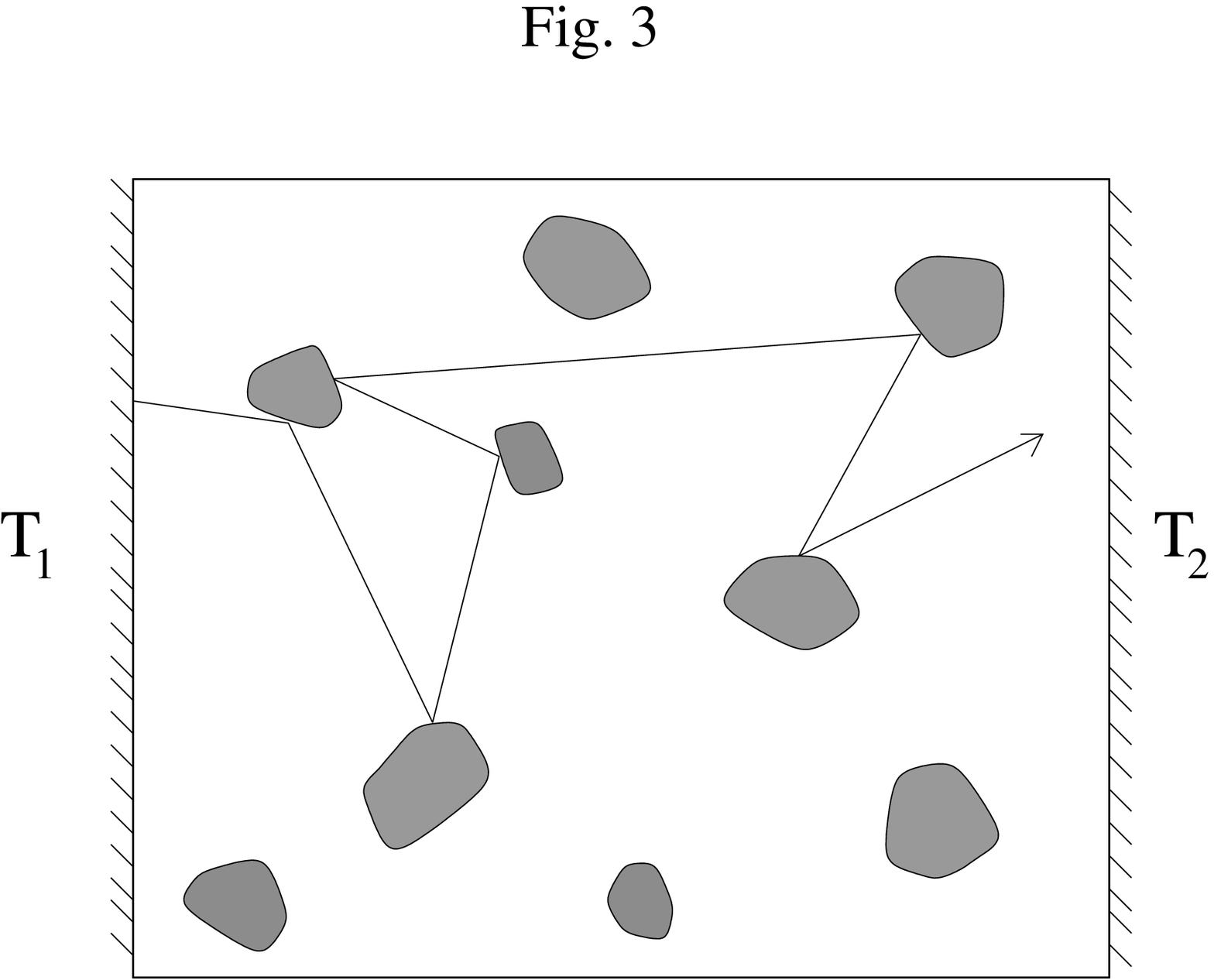}
\begin{figure}
\caption{\label{} 
A schematic representation of the Lorentz model. Particles move
ballistically, and are scattered elastically, but in a random direction on
collision with an obstacle. At the left and right boundaries, they are
reflected and are given a new energy randomly from a distribution
corresponding to two different temperatures $T_1$ and $T_2$ } 
\end{figure}}

\end{document}